\documentclass[a4paper,11pt]{article}
\usepackage{jheppub} 
\usepackage{amsmath, amsfonts, amsthm, bbm}
\usepackage{amsthm}
\usepackage{amssymb}
\usepackage{pifont}
\usepackage{graphicx} 
\usepackage{bmpsize}
\usepackage{braket}
\usepackage{appendix}

\usepackage{physics}
\usepackage{siunitx}

\usepackage{tikz}
\usetikzlibrary{quantikz}

\title{\boldmath Low-temperature Holographic Screens Correspond to Einstein-Rosen Bridges}

\author{Marco Alberto Javarone}

\affiliation{Dipartimento di Fisica, Università di Bari,\\
I-70126 Bari, Italy}
\affiliation{Dutch Institute for Emergent Phenomena,\\ Amsterdam, Netherlands}

\emailAdd{marcojavarone@gmail.com}

\abstract{

Recent conjectures on the complexity of black holes suggest that their evolution manifests in the structural properties of Einstein-Rosen bridges, like the length and volume.
The complexity of black holes relates to the computational complexity of their dual, namely holographic, quantum systems identified via the Gauge\slash Gravity duality framework.
Interestingly, the latter allows us to study the evolution of a black hole as the transformation of a qubit collection performed through a quantum circuit.
In this work, we focus on the complexity of Einstein-Rosen bridges. More in detail, we start with a preliminary discussion about their computational properties, and then we aim to assess whether an Ising-like model could represent their holographic dual.
In this regard, we recall that the Ising model captures essential aspects of complex phenomena such as phase transitions and, in general, is deeply related to information processing systems.
To perform this assessment, which relies on a heuristic model, we attempt to describe the dynamics of information relating to an Einstein-Rosen bridge encoded in a holographic screen in terms of dynamics occurring in a spin lattice at low temperatures. We conclude by discussing our observations and related implications.
}

\begin{document}
\maketitle
\flushbottom

\section{Introduction}\label{sec:introduction}
Eternal black holes~\cite{maldacena01} contain Einstein-Rosen bridges (ERB), i.e. wormholes able to connect very distant universes. Such property makes wormholes particularly fascinating, as reflected in an interest not limited to the scientific community, e.g. plots of some fantasy novels mention ERBs and similar structures. 
These bridges, anchored to entangled surface areas, are likely related to the phenomenon of entanglement, as suggested by the \textit{ER = EPR} conjecture~\cite{susskind06,susskind05}. The latter states that an ERB underlies the connection between entangled particles ---see also~\cite{jensen01,nogueira01}.
In addition, recent investigations~\cite{susskind05,susskind03,susskind09} claim that ERBs are deeply related to the computational complexity (hereinafter just complexity) of black holes. 
As known in Information Theory and related domains, complexity measures quantify the resources required to accomplish a task, e.g. the time an algorithm takes to complete a computation.
Studying black holes via the Gauge\slash Gravity duality framework~\cite{witten01,witten02,ramallo01}, their complexity can be identified by looking at their dual (i.e. holographic) quantum systems.
Accordingly, given two systems dual to each other, e.g. a Qubit collection and a black hole, an increase in complexity in one system is accompanied by an increase in complexity in the other. 
The complexity of the Qubit collection can increase by processing Qubits through a quantum circuit. Instead, the complexity of a black hole is supposed to grow during its evolution. Therefore, a measure of complexity describing the quantum system should have a corresponding (i.e. dual) complexity measure in the Anti-de-Sitter (AdS) space, which must relate to the evolution of the black hole. Among the possible AdS candidates for manifesting such complexity, we find some properties of the ERB, e.g. the ERB length~\cite{susskind03}, the ERB volume~\cite{susskind09}, and the Wheeler-DeWitt (WDW) action~\cite{susskind10,susskind11}. 
Now, the evolution of a physical system occurs during a time interval, typically ranging from a given instant, say $t_0$, to another time $t > t_0$ that, for some systems, may correspond to an equilibrium (or steady) state.
Yet, to study the evolution of black holes, we need to go beyond the classical perspective, which describes them as static objects and find a suitable time interval.
In these objects, the definition of a state of equilibrium is far from trivial thus, to find a meaningful time interval for studying their evolution, we may refer to the scrambling process~\cite{lashkari01}. The latter constitutes a sort of thermalisation process, albeit the two processes are different.
Accordingly, we can study black hole dynamics from their formation to the end of the scrambling process. In this regard, authors of~\cite{susskind10} suggest that, since black holes are the fastest scramblers in the universe, they are also the fastest computers. Such a remarkable claim aligns with the core message of the 'It from Qubit' slogan~\cite{deutsch01} at the base of a fruitful research direction.
In this work, we focus on the complexity of ERBs, aiming to shed light on their nature.
To this end, we start analysing three possible classes of ERBs, defined according to computational properties. Such preliminary analysis, performed to get some insights into the considered system, is followed by the definition of a holographic description of the formation and growth of an ERB. 
The resulting model has a heuristic nature and assumes ERBs are responsible for entanglement~\cite{maldacena02} and transmission of information.
More specifically, the proposed model refers to the dynamics of information encoded in a holographic screen, which is assumed to be composed of square cells as 'pixels'~\cite{verlinde01}, containing information about phenomena and structures manifesting in an AdS space. 
In doing so, we consider mapping the dynamics of correlated patterns emerging in the holographic screen to dynamics occurring in an Ising-like model. Following this idea, we describe the holographic screen as a structure endowed with a bi-dimensional lattice containing spins. Accordingly, in this picture, the spatial spin correlations emerging at low temperatures correspond to the formation of an ERB and support the related entanglement.
Before proceeding, let us remark that assuming the holographic information gets encoded in the mentioned screen does not prevent the existence of additional screens carrying more copies of the same AdS phenomenon, albeit even encoded differently or related to other properties of the same system.
Moreover, going to the behaviour of the spins, arranged in the holographic screen that we study, uncorrelated configurations dominate at high temperatures, under specific conditions~\cite{zhang01,vecsei01}, both in the classical and quantum domain. Likewise, spins form configurations that minimise the free energy at low temperatures. 
Interestingly, the resulting dynamics allow us to relate the complexity of the spin model with the ERB volume, recovering the conjecture proposed in~\cite{susskind09}.
It is worth specifying that our attempt to represent ERB dynamics in terms of an Ising-like model relies on the plethora of phenomena this class of models allows studying, from phase transitions to information processing~\cite{nishimori01}.
In addition, we remark that both scrambling and thermalisation lead to a system transformation, albeit at different time scales. Regarding this point, we show how to face this relevant criticality in the next sections.
In summary, in light of the similarities and differences we identify between the (dual) ERB dynamics and an Ising model, we try to report relevant strengths and weaknesses of the proposed correspondence.
To conclude, the remainder of the work is organised as follows. Section~\ref{sec:bh_complexity} summarises essential concepts of quantum complexity and fundamental conjectures at the base of our model. Section~\ref{sec:complexity_erb} presents some computational properties of ERBs. Then, Section~\ref{sec:heuristic_model} describes our heuristic model.
The manuscript ends with observations and potential developments in Section~\ref{sec:conclusions}.
\section{Complexity and Evolution}\label{sec:bh_complexity}
\begin{figure}[h]  
 \centering
    \includegraphics[width=3.0in]{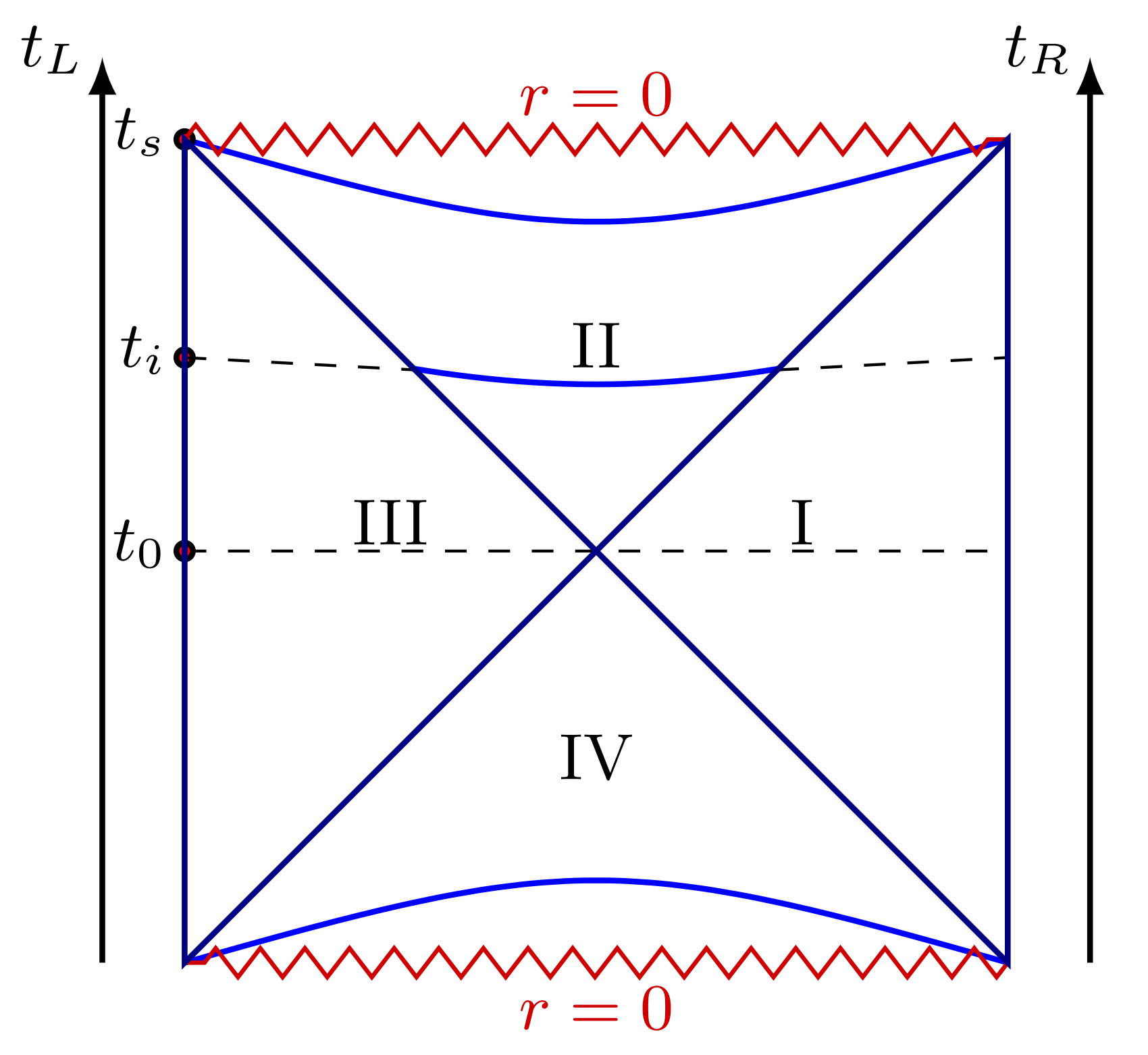}
\caption{AdS eternal black hole. The ERB is identified at time $t_i$, considering a common origin for $t_L$ and $t_R$ at $t_0$.} \label{fig:figure_1}
\end{figure}

In general, complexity quantifies the difficulty of a task, e.g. in performing a transformation. Considering a physical system whose possible states (or configurations) form a phase space, a transformation traces a path in such space and the related complexity is proportional to the length of the path. Taking two points of phase space, the higher the distance between them, the higher the complexity of the transformation connecting them~\cite{javarone02}.
Typically, in spontaneous transformations, paths correspond to geodesics between the initial and the final state.
Also, the complexity of a transformation corresponds to the complexity of the final state, defined by taking as a reference the initial state. In this sense, complexity is a relative measure which allows us to compare more states by taking one as a reference ---see ~\cite{susskind05,susskind03,haferkamp01} for additional details.
\newline
These considerations also apply to black holes, namely their evolution must be accompanied by a growth of their complexity.
Yet, computing the complexity of a black hole is far from trivial. Notwithstanding, a series of recent works (e.g.~\cite{susskind03,susskind09,susskind12}) that rely on the gauge/gravity duality framework identify fascinating correspondences to perform this calculation. 
More in detail, these approaches consider quantum systems, described by state vectors, dual to black holes whose transformations can be realised via generic operators, e.g. quantum gates.
For instance, a state vector $\ket{phi}$ representing a system with $K$ Qubits evolves under the time operator $U(t) = e^{-iHt}$, i.e. $\ket{\phi_t} = U\ket{\phi_0}$.
In turn, the operator $U$ can be decomposed into more simple operators, as $U = g_x...g_1$. Each simple operator $g_i$ belongs to the collection of operators ${G}$, which we can use to build $U$, and may correspond to a quantum gate. 
For clarity, if an operator $O_T$ can be decomposed as $O_T = o_1 \cdot o_0$, then we assume that $o_0$ and $o_1$ are simpler than $O_T$.
Given that $U$ is a unitary transformation, the number of operators $g_i$ we can use to build $U$ has an upper bound. Accordingly, under this operator, a state vector $\ket{\phi}$ has an upper bound in the complexity (i.e. $C_{max}$).
By mapping operators to quantum gates, we can study quantum circuits able to transform collections of Qubits and measure the related growth in complexity. 
Hence, given a system with $K$ Qubits, a quantum circuit has a computation rate $C_r = K/\Delta t$. This rate corresponds to the number of quantum gates that can be processed in one interval of time (that can also be expressed as a thermal time $\Delta t = \frac{1}{T}$). 
Thus, after a time interval of length $t$, the complexity of the final quantum state reads
\begin{equation}\label{eq:complexity_and_rate}
C(t) = C_r t.
\end{equation}
\noindent A fundamental question addressed by Lloyd~\cite{lloyd01}, and adapted to the specific context in~\cite{susskind10}, relates to the limits of computation of physical systems. These limits can, for instance, interest storage, resources, and processing units. 
For a quantum circuit with an energy $E$, there is a limit in the number of operations (i.e. gates) per second equal to 
\begin{equation}\label{eq:lloyd}
\frac{d \text{ NrGates}}{dt} \le \frac{2E}{\pi \hbar}.
\end{equation}
\noindent Interestingly, equation~\ref{eq:lloyd} can be used even for studying the evolution of a quantum system dual to an eternal AdS-Schwarzschild black hole. 
To this end, we take two instants of time $t_L$ and $t_R$, writing the state $\ket{\phi(t_L,t_R)}$ as
\begin{equation}\label{eq:state_ads_schw_bh}
    \ket{\phi(t_L,t_R)} = e^{-i(H_Lt_L + H_Rt_R)}\ket{\text{TFD}}
\end{equation}
\noindent so that this state originates from $\ket{\text{TFD}}$ defined as 
\begin{equation}\label{eq:state_ads_schw_bh}
    \ket{\text{TFD}} = Z^{-1/2}\sum_{\alpha} e^{-\frac{\beta E_\alpha}{2}}\ket{E_{\alpha}}_L\ket{E_{\alpha}}_R
\end{equation}
\noindent where $Z$ denotes the partition function, and $TFD$ denotes the Thermofield-Double State (which is convenient to have two identical quantum mechanical systems~\cite{jefferson01,cottrell01}).
In this system, the time evolution operator $e^{-i(H_Lt_L + H_Rt_R)}$, in function of $t_L$ and $t_R$ (see Figure~\ref{fig:figure_1}), allows $\ket{\text{TFD}}$ to evolve and such evolution gets reflected in the complexity of $\ket{\phi(t_L,t_R)}$.
Eventually, equation~\ref{eq:complexity_and_rate} and the bound~\ref{eq:lloyd} allow us to identify a black hole quantity dual to the complexity of the holographic quantum system. For more details see~\cite{jefferson01,chapman01,khan01}.
\subsection*{Complexity of Black Holes}
Recent works~\cite{susskind03,susskind09,susskind10,susskind11,susskind12} present various correspondences for identifying the complexity of black holes. In these investigations, authors assume that at $t=0$, i.e. at the instant of formation, black holes are in the simplest possible state. 
The simplest quantum state corresponding to the newly formed black hole is $\ket{\phi(t_0)} = \ket{000\dots0}$.
A suitable time interval for studying the evolution of the black hole goes from its formation ($t_0$) to the end of the scrambling process ($t_s$).
Looking at the dual quantum system composed of $K$ Qubits, once scrambling is completed, the complexity is $C(t_{s}) = K \log K$.
Assuming $K$ is equal to $S$, i.e. the number of degrees of freedom of a black hole (namely its entropy), the complexity of the quantum system is $C(t_{s}) = S \log S$.
Notice that, according to equation~\ref{eq:complexity_and_rate}, the complexity grows linearly~\cite{haferkamp01} as $C(t) = STt$, and this trend can be approximated to
\begin{equation}\label{eq:complexity_low_t}
C(t) \sim St
\end{equation}
\noindent for low temperatures.
These relationships, as briefly presented here, allow us to compute the complexity of black holes.
In~\cite{susskind09}, authors propose a conjecture named $C = V$, i.e. 'Complexity equals Volume' of the ERB, deriving the following equation
\begin{equation}\label{eq:cv}
C = \frac{V}{l}
\end{equation}
\noindent with $l$ corresponding to some radius, such as the AdS radius and Schwarzschild radius.
Here, we report only a few equations for connecting later our model to this discussion. In addition, from now on, if not stated otherwise, quantities are written adopting the natural units convention so that $G= \hbar = c = 1$.
Let us begin with the volume of the ERB $V$ that, in a Schwarzschild metric, reads
\begin{equation}\label{eq:schw}
ds^2 = -f(r)d\tau^2 + f(r)^{-1}dr^2 + r^2 d\Omega_{D-2}^2
\end{equation}
\noindent with
\begin{equation}\label{eq:fr}
f(r) = r^2 + 1 - \frac{16 \pi M}{(D-2)\omega_{D-2} r^{D-3}}
\end{equation}
\noindent and $M$ mass of the black hole. As shown in~\cite{susskind09}, the temporal variation of $V$ is equal to
\begin{equation}\label{eq:temp_v}
\frac{dV}{d\tau} = \omega_{D-2}r^{D-2}\sqrt{|f(r)|}
\end{equation}
\noindent and the above quantity is maximised at a value of $r$ defined as $r_m$, identified by a term $v_D = \omega_{D-2}r^{D-2}\sqrt{|f(r_m)|}$. 
Then, the volume $V$ becomes
\begin{equation}\label{eq:volume}
V = v_D \Delta t
\end{equation}
\noindent with an interval of time, $\Delta t$, having a spacelike behaviour since related to the time 'inside' the black hole.
Given the time variables $t_R$ and $t_L$, related to the two universes $I$ and $III$, respectively, the above time interval is $\Delta t = |t_R + t_L|$.
Eventually, the volume of the ERB connecting the sectors $I$ and $III$ in Figure~\ref{fig:figure_1} reads
\begin{equation}\label{eq:v_erb}
V = v_D |t_R + t_L|.
\end{equation}
The $v_D$ term contains the connection between the complexity of the dual quantum system and the ERB volume. In some conditions its value is $v_D = \frac{8\pi M l}{D-2}$. 
Since $M \sim ST$, i.e. the mass of the black hole is proportional to the product between its entropy $S$ and temperature $T$, the correspondence $C = V$ (see equation~\ref{eq:cv}) is realised by equation~\ref{eq:v_erb} as one can see comparing with equation~\ref{eq:complexity_and_rate} and equation~\ref{eq:complexity_low_t}.
A refinement to this correspondence led to the conjecture $C = A$, i.e. 'Complexity equals Action', introduced to reduce, as much as possible, the arbitrariness adopted for deriving the correspondences between complexity and physical quantities such as volumes or lengths.
Finally, let us report a brief comment about the conjecture $C=A$. The latter is interesting as, in classical mechanics, the action is highly related to a transformation (or an evolution).
For example, a classical action $A_{c}$ for a simple Lagrangian $L$, composed only of a kinetic term $T$, is equivalent to $A_{c} = \int dt L = \int dt T$.
So, a complexity growth corresponding to the evolution of a system described by the Lagrangian $L$ can intuitively be associated with $\frac{A_{c}}{dt}$, since $\frac{dA_{c}}{dt} = T$.
In light of this observation, the bound in equation~\ref{eq:lloyd} could find its roots in foundational relationships of mechanics.
\section{Complexity of Einstein-Rosen bridges}\label{sec:complexity_erb}
In this section, we present some observations on the computational complexity of Einstein-Rosen bridges.
Let us recall that these structures are supposed to connect distant locations, and their surface areas are entangled. Accordingly, information should get transmitted through them. 
Now, since the Gauge/Gravity duality plays a fundamental role in computing the complexity of objects in the AdS spacetime, we need to identify a suitable dual quantum system for an ERB. Such a dual system, whose evolution must reflect the growth of the associated structure (i.e. the ERB), has to support the transmission of information, i.e. to work like a circuit or communication channel.
\begin{figure}[h]  
 \centering
    \includegraphics[width=3.0in]{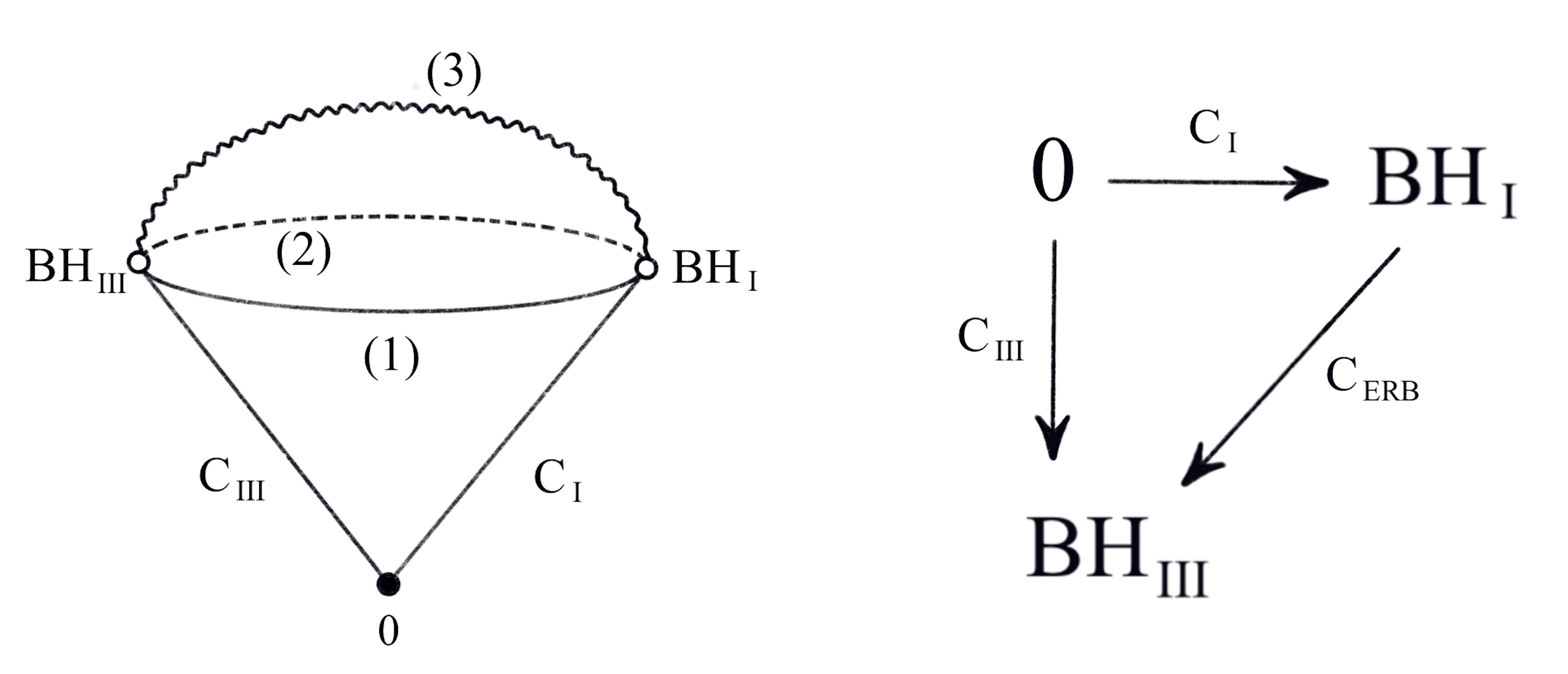}
\caption{Pictorial image of an eternal black hole evolving from the origin in $0$ to two entangled areas $BH_{I}$ and $BH_{III}$. Three different paths represent ERBs with different complexity (with relation to $0$): decreasing complexity $(1)$, equal complexity $(2)$, and increasing complexity $(3)$. The transformation implementing the path from $BH_{I}$ to $BH_{III}$ is shown in the diagram on the right. Details are discussed in the main text.} \label{fig:figure_2}
\end{figure}
Before proceedings, let us observe that, in principle, any path in an AdS space could have a dual quantum system. Yet, exceptions may show up for those paths that cannot be traversed by particles, e.g. due to causal structure constraints. 
Studying whether a dual system can or cannot exist is beyond the scope of this work. Here, we consider only paths related to ERBs ---see the diagram in figure~\ref{fig:figure_2}, i.e. paths corresponding to quantum systems connecting the configuration a black hole area has in the universe $I$ with that a black hole area has in the universe $III$. Then, as mentioned, these quantum systems are represented as circuits.
\subsection{Complexity of ERB Circuits}
As mentioned before, computing the complexity of a state requires identifying a reference state~\cite{susskind02}.
So, taking as reference the Qubit configuration dual to an eternal black hole at the instant of formation (i.e. $t = t_0$), we can classify an ERB according to the following classes:
\begin{enumerate}
    \item Paths of decreasing complexity;
    \item Paths whose complexity remains invariant;
    \item Paths of increasing complexity.
\end{enumerate}
Note that at $t > 0$, from the point of view of the black hole surface in the universe $I$, the ERB corresponds to a circuit whose complexity is always $C > 0$. That is motivated once we relate the complexity of a path with its length. 
Also, in mapping paths to circuits, a path with no complexity (i.e. $C = 0$) corresponds to a circuit composed of buffers, as the input is equivalent to the output.
Then, in light of these three classes, we aim to find the most suitable one for an ERB able to support the entanglement.
\subsubsection{ERB case $(1)$}
In geometrical terms, an ERB path of decreasing complexity corresponds to a path going towards the black hole's origin (i.e. at $t = t_0$). Thus, assuming that the back hole evolution occurs only along the time direction, getting closer to its origin is equivalent to travelling back in time.
Remarkably, particles as electrons can do that~\cite{banerjee01}. Consequently, a quantum observer could, in principle, accept such an implication.
Let us identify the following quantum circuits: $c_{erb}$, $c_I$, and $c_{III}$. The first one, i.e. $c_{erb}$, corresponds to an ERB. The two additional circuits correspond to the paths from the black hole origin (at $t_0$) to the black holes in the universe $I$ and $III$, respectively. All these circuits are reversible~\cite{nielsen01,wille01}, so we can write the following equivalences:
\begin{align*} 
c_{III} = c_{erb} \circ c_I\\
c_{III} \circ c_I^{-1} = c_{erb} \circ c_{I} \circ c_{I}^{-1}\\
c_{erb} = c_{III} \circ c_{I}^{-1}
\end{align*}
\noindent where $c_I^{-1}$ denotes the inverse circuit $c_I$. 
Therefore, as $\ket{\phi_0} = c_{I}^{-1}(\ket{\phi_{t}})$, with $\ket{\phi}$ representing the Qubit collection dual to a black hole, an ERB of decreasing complexity entails travelling back to the origin and then evolving to the opposite equivalent black hole in the other universe.
\subsubsection{ERB case $(2)$}
Here, we consider an ERB path whose complexity does not change over time, namely $\frac{dC}{dt} = 0$. We recall that, inside the black hole, time has a space-like nature, and the complexity is computed by taking as reference the state of black hole formation.
The circuit dual to such path generates a sequence of quantum configurations having the same complexity as those corresponding to the black holes in universes $I$ and $III$, respectively. Indeed, the input, output and intermediate configurations along the path do not need to be the same. Yet, these have the same complexity in relation to the state of reference. To clarify this point, black hole configurations can be thought of as leaves of an evolutionary tree, so having the same complexity from the root (i.e. black hole at $t=t_0$) entails being at same the level in the tree~\cite{javarone01}. 
In geometrical terms, this path goes along an arc of the circumference at a fixed distance from the origin ---see~\ref{fig:figure_2}.
The black hole complexity, as computed in~\cite{susskind09}, is proportional to $C \sim (t_L + t_R) \cdot S$. 
In the middle of this hypothetical ERB, assuming $t = t_L = t_R$, the complexity is about $C_{\frac{1}{2}ERB} \sim t \cdot S$, with $t$ the time the black hole has evolved since its formation (by setting $t_0 = 0$).
Moreover, the 'external observer' located in the universe $I$ (or $III$) would measure an increase of complexity along the ERB, though along a path which is not geodesic towards the entangled black hole. As we mentioned, the ERB would be along the arc of circumference connecting the two black hole areas, so the rope between them is shorter than such an arc. Therefore, assuming the principle of minimum action holds in this physical system, no spontaneous evolution may occur along an ERB whose complexity does not change from the point of view of the observer located at the origin.
To conclude, this path may exist even if, given the reason above-mentioned and further considerations below discussed, we discard this possibility.
\subsubsection{ERB case $(3)$}
Finally, before considering an ERB path of increasing complexity, we recall that the $C=V$ conjecture~\cite{susskind09} states that the increase in the black hole complexity is proportional to the growth of its ERB. 
Thus, a constant surface area of the black hole implies the ERB becomes longer. Consequently, a quantum circuit dual to this special ERB is a circuit of increasing length. 
For instance, we can imagine a circuit whose amount of gates increases over time. 
Interestingly, this scenario can be described as a sort of thermodynamic limit for getting out from such an ERB due to a continuously growing length. 
In addition, after a huge time interval, the configuration processed by this quantum circuit would reach a maximum value so, as described in~\cite{susskind02}, its complexity would drop down. The time scale for completing this process is proportional to $e^{e^S}$ (where $S$ denotes the black hole entropy).
To conclude, a thermodynamic limit in the traversability of an ERB would not be too surprising. However, we discard this option since the time scales for connecting the two areas are so high that the path could be similar to the one connecting universes $I$ and $III$ travelling outside the black hole.
\newline
\newline
\noindent After discussing these three possibilities, we suggest that the case $(1)$, i.e. paths of decreasing complexity constitutes the more suitable option for representing an ERB. Also, we remark that paths $(2)$ and $(3)$ are affected by an additional physical constraint, i.e. they would cross the coordinate $r=0$ shown in figure~\ref{fig:figure_1} and, in addition, would entail travelling in the future and then going back to the present time. 
Eventually, in agreement with the above observations, we propose a heuristic model to describe the emergence of entangled surface areas in the AdS space from an eternal black hole. Our model shows the formation of an ERB whose structure cannot be preserved forever, i.e. it has a temporary nature, in line with the phenomenon of entanglement. For these reasons, the proposed model is complementary to that proposed in~\cite{susskind09}, which refers to a quantum circuit. 
Notice that the proposed model aims to represent the dynamics of information encoded in one holographic screen located at the boundary of an AdS space, among the others which contain information about AdS Physics.
\section{Towards a holographic ERB model}\label{sec:heuristic_model}
As described before, according to ~\cite{susskind03,susskind09,susskind10,susskind12}, the complexity of a black hole increases while scrambling. That seems intuitive in light of~\cite{nielsen02}, considering that this process entails a transformation. Notably, any transformation changing (the state vector of) a quantum system implies a variation in complexity.
Hence, if a quantum system constitutes the holographic dual of a black hole, any complexity growth in the former leads to a complexity growth in the latter and vice versa. 
Moreover, the scrambling of a black hole, from the point of view of a dual Qubit collection starting in the state $\ket{\phi} = \ket{000...0}$, resembles an order-disorder phase transition from an ordered configuration to a disordered one. 
Similarly, in some spin systems, such as the Ising model with ferromagnetic interactions, the above transition occurs when the temperature initially low heats up to values higher than the critical one ($T_c$).
For clarity, scrambling is not an order-disorder phase transition. Yet, considering a dual Qubit collection, the latter and the Ising model may describe different properties of the same phenomenon.
\begin{figure}[ht!]
    \centering
    \includegraphics[width=6.0in]{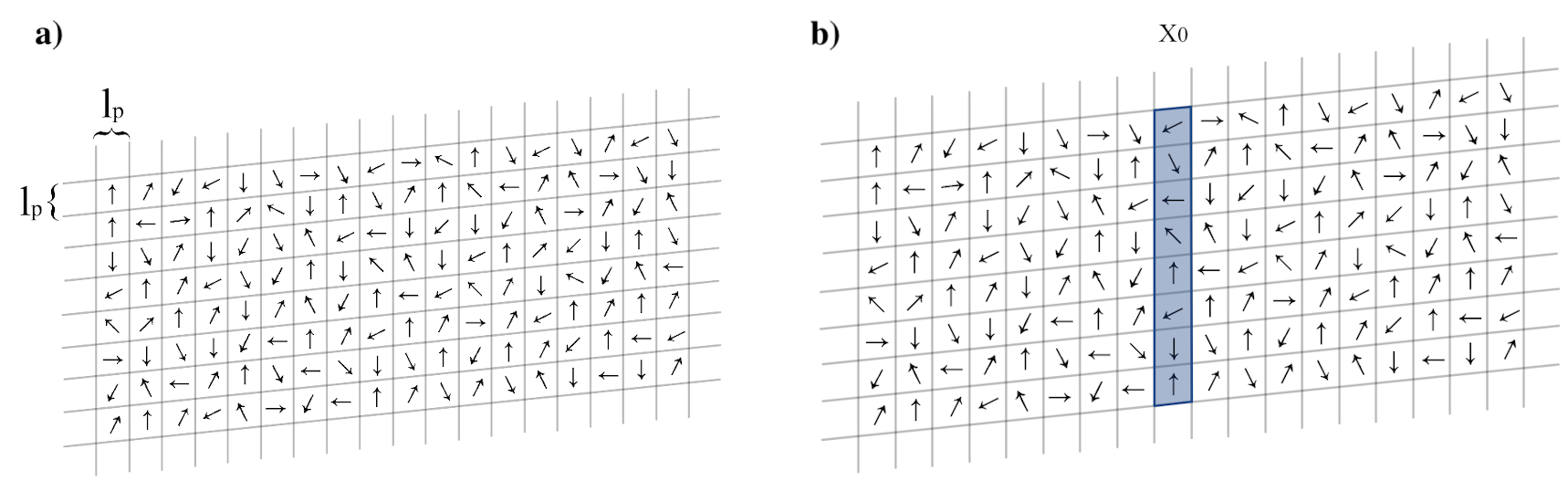}
    \caption{ \textbf{a)} Lattice organisation of the holographic screen. Each cell contains a spin and has size $l_p \times \l_p$. \textbf{b)} The formation of an eternal black hole is dual to the collection of spins (highlighted over the blue column) in the holographic screen. The spatial coordinate $x_0$ allows us to identify the collection of spins dual to the event horizon of the black hole at the instant of formation.}
    \label{fig:figure_3}
\end{figure}
More specifically, scrambling is related to a transformation occurring in the dual quantum system, leading simple configurations to become (more) complex configurations. Also, according to the central dogma of black holes, it takes place inside them.
For that reason, in the time interval going from black hole formation $t_0$ to the end of the scrambling $t_s$, we suggest that a phenomenon similar to order-disorder phase transition takes place on a holographic screen.
In this regard, we propose an Ising-like model with a complexity equivalent to that computed in~\cite{susskind09}, whose spin correlation may be related to the entanglement between surfaces anchored to an ERB. 
We finally remark that the proposed model agrees with the hypothesis $(1)$, previously discussed, concerning the complexity classes of ERBs.
\subsection*{Structure of the Holographic Screen}
\begin{figure}[ht!]
    \centering
    \includegraphics[width=6.0in]{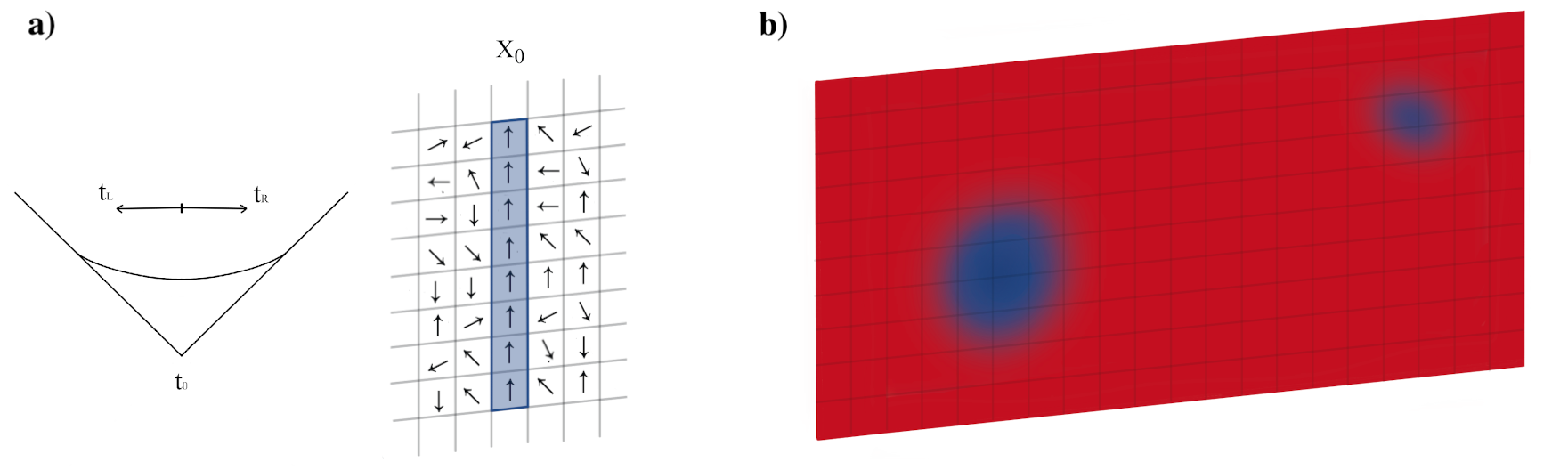}
    \caption{ \textbf{a)} The evolution of an eternal black hole identified in a holographic screen, whose spins correlated in both directions starting from the pattern at $x_0$. \textbf{b)} Heatmap representing a pictorial image of the holographic screen. The red colour indicates high temperatures, while the blue areas correspond to portions of the screen at low temperatures.}
    \label{fig:figure_4}
\end{figure}

As hypothesised in~\cite{verlinde01}, a holographic screen contains information related to the physics occurring in an AdS space. The holographic encoding is redundant and non-local, so building a whole system requires more screens.
Here, we focus on the properties of a single screen, assuming it contains only the information related to some properties of an ERB as those to support the black hole entanglement.
The considered screen is composed of cells, as pixels, each of size $l_p \cdot l_p$ ($l_p$ Planck length), and contains an amount of information corresponding to that of a spin particle $\sigma$. For simplicity, we place a spin on each cell of the screen.
The emerging structure resembles a bi-dimensional spin-lattice, corresponding to an Ising model in dimension $D=2$ ---see panel \textbf{a)} in Figure~\ref{fig:figure_3}. We assume the screen can be described as a thermodynamic system, so we identify variables such as temperature.
In agreement with that, spins fluctuate and form various patterns. Here, a holographic screen with ferromagnetic interactions and a non-uniform temperature can show clusters of aligned spins. Yet, an overall high temperature would result in a global spin distribution showing mostly disordered patterns.
Including non-ferromagnetic interactions, ordered patterns cannot form. Yet, at low temperatures, those patterns minimising the free energy of the system show up.
In this picture, we consider a subset of spins organised into a column on the screen whose interactions are ferromagnetic. These spins constitute the dual of an AdS eternal black hole and, for being congruent with the dynamics of the Qubit collection in~\cite{susskind09}, at $t=t_0$, they are in the state of minimum energy (e.g. all spins equal $\sigma = +1$) ---see Figure~\ref{fig:figure_4}.
At this point, to observe correlated spins across the screen, the temperature must be lower than the critical one $T_c$. If that condition is non-permanent, the black hole configuration and the dual ERB we are building cannot last forever. That seems coherent with the non-permanent nature of entanglement (see e.g.~\cite{lin01}). 
Following this description, the dynamics of the information encoded as spin patterns, dual to the ERB, can be studied via an Ising model. 
The latter, for clarity, here includes both ferromagnetic and non-ferromagnetic interactions.
\subsection*{Spin Lattice - ERB Duality}
We consider an Ising model with the following Hamiltonian
\begin{equation}\label{eq:hamiltonian}
    H = - \sum_{i,j}^{N,N} J \sigma_i \sigma_j
\end{equation}
\noindent where $J$ denotes the interaction term between spins $\sigma_i$ and $\sigma_j$. For simplicity, in equation~\ref{eq:hamiltonian}, additional terms such as external fields are omitted.
The holographic dual of the black hole, i.e. the collection of spins, is identified via a spatial coordinate ($x_0$) on the screen. Thus, starting from $x_0$, at low temperatures, a growing set of correlated spins spreading on both sides of the screen emerges ---see panel~\textbf{b} of Figure~\ref{fig:figure_3}). 
Also, as the temperature approaches the critical value $T_c$, the correlation length $\zeta$ grows till diverging at infinity, following a behaviour described by 
\begin{equation}\label{eq:correlation_length}
    \zeta = |\tau|^{-\nu}
\end{equation}
\noindent where $\nu$ denotes the critical exponent and a $\tau$ dimensionless parameter defined $\tau = \frac{T - T_c}{T_c}$.
The spatial correlation between spin pairs reads
\begin{equation}\label{eq:spatial_correlation}
    \langle\sigma_0 \sigma_x \rangle \sim e^{-\frac{x}{\zeta}}
\end{equation}
\noindent where $x$ represents the spatial distance on the lattice between spins.
We recall that cells, or pixels, which contain spins have a size proportional to $l_p$, equal to a discrete time interval. 
Therefore, the amount of time the (column of) spins needs to correlate in both directions can be derived by dividing the maximum spatial distance reached at time $t$, say $L_{max}$ obtained via equation~\ref{eq:spatial_correlation} by $l_p$, so that
\begin{equation}\label{eq:time_int}
t_{max} = L_{max} / l_p
\end{equation}
\noindent where $L_{max}$ is obtained for the highest $x$ such that $\langle\sigma_0 \sigma_x \rangle \sim 1$.
Following the above description, we argue that the spatial correlation of spins propagating through the two sides of a screen corresponds to the formation of the ERB volume. As defined in equation~\ref{eq:v_erb}, this volume results from the surface entropy $S$ correlated with an increasing amount of patterns (in both directions, i.e. left and right) over a time of duration $|t_L + t_R|$.
At this point, there are some aspects to clarify. Firstly, we assume the temperature fluctuation on the screen involves an increasing number of spins. 
That is possible by a fluctuation spreading at a given speed over the screen. That can fix a fundamental issue, i.e. thermalisation is faster than scrambling. The former occurs within a time proportional to $\frac{1}{T}$~\cite{jefferson01}. However, here, thermalisation takes place on an increasing spatial scale. In doing so, uncorrelated spins on the lattice thermalise to correlated patterns as soon as the portion of the screen where they are localised is affected by the temperature fluctuation.
For clarity, we assume that the mentioned temperature fluctuation, responsible for reducing the average screen temperature, spreads according to a time scale compatible with the time scale of scrambling. Hence, after that amount of time, high temperatures restore the thermodynamic conditions on the screen, and the spin correlations vanish.
Eventually, assuming a quantum Ising-like model on the holographic screen, the entanglement entropy can be quantified by a universal formula~\cite{calabrese01}:
\begin{equation}\label{eq:entanglement}
EE = \frac{q}{3}\log (\frac{\zeta}{l_p})
\end{equation}
\noindent that applies to one-dimensional spin chains.
In this regard, note that by considering only the entanglement between pairs of spins forming the two entangled black holes~\cite{susskind_tele}, i.e. that in the universe $I$ and that in $III$, equation~\ref{eq:entanglement} might also work for the system we are considering. 
Following the above assumption and elaborating equation~\ref{eq:entanglement} with some fixed numerical values ($q = l_p = 1$, where $q$ denotes the central charge), we get an entanglement entropy proportional to the logarithm of $\zeta$, i.e. $EE = \frac{1}{3} \log(\zeta)$, such that $EE > 0$ at low temperatures. 
In doing so, if each pattern on the screen corresponds, i.e. constitutes the holographic dual, to a black hole surface, according to the described Ising-like mechanism an 'external observer' can measure non-local correlations.
\subsection*{Complexity of the Dual Spin Model}
As reported, complexity quantifies the minimum number of transformations required for obtaining a specific state or configuration starting from one of reference. In our case, the latter lies on the screen and has spatial coordinate $x_0$.
So, we can calculate the number of transformations occurring from $x_0$ to $L_{max}$ on both sides due to the spreading of the temperature fluctuation that reduces the screen temperature locally. 
The column pattern located at $L_{max}$ contains the sequence of spins $\{\sigma_{0,L_{max}},...,\sigma_{S,L_{max}}\}$, whose indices denote the spatial coordinates of cells in the lattice.
The spin collection at $L_{max}$, initially in a random configuration, fluctuates from one state to another. All these variations are not related to the final sequence. However, once the low-temperature fluctuation gets closer and the nearest spin column correlates with that at the origin $x_0$, spins at $L_{max}$ align according to the related interactions.
Here, calculating the number of transformations is recursive and leads to a complexity equal to
\begin{equation}\label{eq:nr_gates}
C = 2S \cdot t_{max}.
\end{equation}
\noindent The factor $2$ in the above equation~\ref{eq:nr_gates} reflects the complexity or the number of transformations required to cover both sides of the screen from $x_0$ to $L_{max}$.
According to the conjecture $C = V$, the ERB volume reads $V = S |t_L + t_R|$. Thus, we obtain a perfect match with $V$ by setting $|t_L| = |t_R| = t(T;v_T)$. More in detail, we find that equation~\ref{eq:nr_gates} is equivalent with equation~\ref{eq:complexity_low_t}.
The time $t(T;v_T)$ is related to the number of patterns correlating with the original one. In addition,  $t(T;v_T)$ depends on the screen temperature (or of a portion of it) $T$ that is reached by the spreading of the fluctuation at a speed $v_T$. The closer the temperature to the critical value $T_c$, the longer the time interval and then the higher the volume of the ERB.
At a specific instant and temperature, the time $|t_L|+|t_R|$ reflects the length spanned by the spatial correlation from the beginning of the process. Also, as mentioned, the low temperature does not remain fixed and tends to increase hence, at some point, the spatial correlation reduces. That implies the entanglement between the two surface areas vanishes as well as the ERB.
\begin{figure}[ht!]
    \centering
    \includegraphics[width=3.5in]{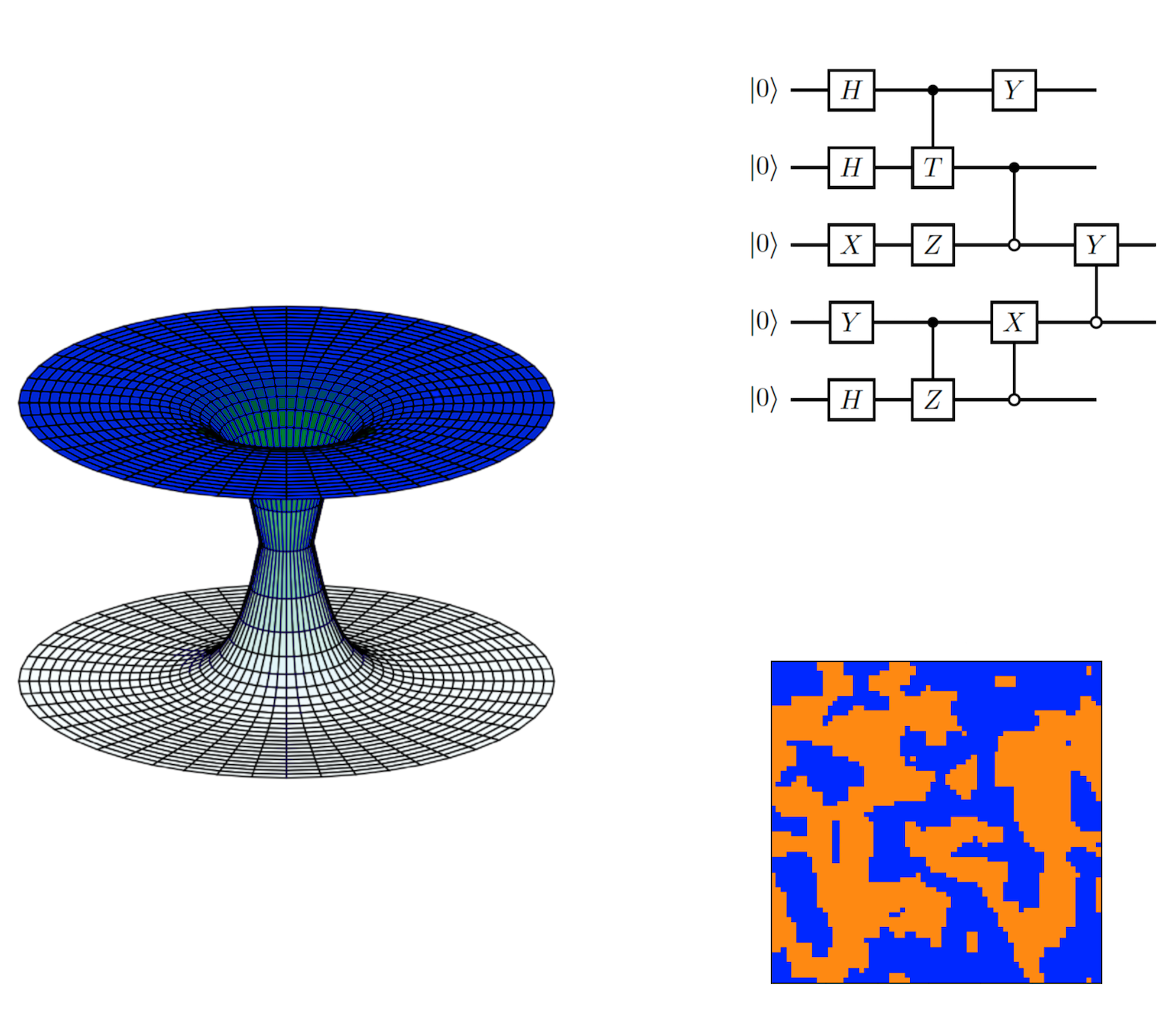}
    \caption{This pictorial image includes two representations of an Einstein-Rosen bridge. Namely, a quantum circuit processing Qubits, and an Ising-like model processing patterns of spins.}
    \label{fig:figure_5}
\end{figure}
These special paths would allow travelling between very distant universes in an extremely brief time. That consideration holds for an 'external observer' whose spatial length of the entanglement encodes in the spatial correlation on the holographic spin lattice.
An 'inner observer' (i.e. inside the black hole) has a time interval equal to the spatial correlation divided by $l_p$, whereas an 'external observer' only measures, in one instant of time, a spatial correlation up to a distance $2 L_{max}$.
To summarise, the complexity of the spin model on the holographic screen equals the volume of the ERB (once parameters are converted as described).
In particular, the whole complexity grows as $C(t) = S \cdot t$, i.e. from $t_0$ until the maximum $t$ allowed by the screen temperature fluctuation.
Eventually, in~\cite{susskind10}, authors state that when a black hole forms, 'it does not begin its life in a random state'. Similarly, we start with a simple ordered spin configuration assuming that, in that area, the temperature is much lower than the $T_c$. 
To conclude, we highlight that the proposed model performs a thermalisation beyond describing the emergence of a spatial correlation dual to the ERB entanglement. A thermalisation occurs as spin columns close to the one in $x_0$ tend to correlate with it (according to the sign of interactions) by undergoing a process that conceptually resembles scrambling, i.e. expected to happen in the AdS black hole. Also, looking at the correlated patterns emerging during the process, we observe a transition from order, identified at $x_0$, to disorder, then recovering the Qubit dynamics described in~\cite{susskind10}.
\section{Discussion and Conclusion}\label{sec:conclusions}
The evolution of eternal black holes is associated with the emergence of Einstein-Rosen bridges that, according to recent investigations~\cite{susskind03,susskind09}, reflect the complexity of these massive objects.
More in detail, the growth of complexity~\cite{susskind02} in black holes, identified via the Gauge\slash Gravity duality, describes their evolution.
As conjectured in~\cite{susskind06, jensen01}, ERBs could be responsible for entanglement at a more general level, e.g. between two particles. Therefore, understanding the holographic conditions and mechanisms underlying their formation and growth is of paramount relevance.
Here, we focus on the computational properties of ERBs. To this end, we start with a preliminary analysis for studying the complexity of paths connecting two entangled black holes.
Resulting observations suggest that these paths should have a decreasing complexity (with reference to the black hole origin).
Then, in accordance with that and the related implications, we propose a heuristic model for the formation and evolution of an ERB. More specifically, our model represents the dynamics of information on a holographic screen.
Before proceeding, we highlight that ERBs of decreasing complexity should have a holographic dual able to process and restore the configurations of a state vector, resembling the dynamics of travelling backwards in time~\cite{banerjee01}. Moreover, we recall that models exploiting quantum circuits may benefit from the fact quantum gates are reversible~\cite{nielsen01,wille01}, albeit that is far from trivial.
Thus, to address the described phenomenon, we study the dynamics of an Ising-like model assuming a collection of spins arranged in the cells which compose a holographic screen. In doing so, the overall system resembles a bi-dimensional lattice, whose dynamics model information is encoded in a screen.
Then, we define some conditions, such as specific thermodynamic properties, to let the spin model evolve at low and high temperatures, forming correlated patterns of various sizes.
Interestingly, we find that the complexity of the process underlying the formation of the correlated spin patterns is the same as the volume of the ERB. That connects our work with previous investigations as~\cite{susskind09}.
Also, the model dynamics can be related to those described in~\cite{verlinde01}, suggesting that the information on holographic screens reflects into the AdS space and may act as memory storage.
Notwithstanding, our proposal has some points requiring clarification. To cite a few, the mechanism underlying the screen temperature, especially its fluctuations and how these propagate, the topological spin organisation, and the distribution of interactions. Shortly, additional work is mandatory to corroborate the validity of our model that, as shown in figure~\ref{fig:figure_5}, can be seen as complementary with that proposed in~\cite {susskind03,susskind09}.
In addition, we remark on the connection of the dynamics here illustrated with entanglement since, as suggested in~\cite{susskind06,jensen01}, this phenomenon could rely on wormholes. More specifically, the proposed model could connect entanglement with low-temperature holographic screens. Yet, as above-mentioned, these observations and speculations require further attention.
Before concluding, we recall one of our initial questions in this work relates to the suitability of an Ising-like model for describing a dual holographic system of an ERB. 
Beyond any required additional investigation, some observations suggest a positive answer. 
To motivate the above comment, we highlight the following points. 
Firstly, we emphasise that the complexity of the proposed model, i.e. equation~\ref{eq:nr_gates}, corresponds to the complexity in equation~\ref{eq:complexity_low_t} and are both identified at low temperatures. That confirms our model agrees with the $C=V$ conjecture.
Going further, the emerging correlation in the Ising model may reflect the emerging entanglement in the eternal black hole, both with a temporary nature (i.e. cannot last forever).
Then, although scrambling and Ising thermalisation are different processes and occur at different time scales, they resemble each other. Also, here we try to fix this issue by assuming a temperature fluctuation spreading and lasting for an amount of time compatible with the time scale of scrambling.
Eventually, the Qubit transformation described in~\cite{susskind10} resembles an order-disorder phase transition. In our model, we can obtain a similar outcome by constraining positive interactions between spins at $x_0$ while allowing both ferromagnetic and non-ferromagnetic interactions among all the other screen cells. More specifically, while the spin pattern in $x_0$ is ordered, correlated patterns, showing up (at low temperatures) on both sides, are disordered due to the mixture of positive and negative interactions.
We conclude by mentioning the dualities between some Ising models and gravity, reported in~\cite{castro01}, that corroborate the relevance of mappings like that we propose.
In summary, in light of the above observations and limits, we deem our model deserves further attention, as it captures relevant aspects of ERBs and could stimulate novel ideas in this direction. 
\acknowledgments
MAJ wishes to thank Sebastian De Haro for his useful comments, Jay Armas for his advice, and Dominik Neuenfeld for the interesting discussions. The author is supported by the PNRR NQST (Code: $PE23$).

\end{document}